# Optimizing Deep Learning Models For Raspberry Pi

Salem Ameen
School of Science, Engineering and Environment
*University of Salford*
Manchester, United Kindom
s.a.ameen1@salford.ac.uk

Kangaranmulle Siriwardana
School of Science, Engineering and Environment
*University of Salford*
Manchester, United Kindom
k.p.s.k.siriwardana@edu.salford.ac.uk

Theo Theodoridis
School of Science, Engineering and Environment
*University of Salford*
Manchester, United Kindom
t.theodoridis@salford.ac.uk

*Abstract*— Deep learning models have become increasingly popular for a wide range of applications, including computer vision, natural language processing, and speech recognition. However, these models typically require large amounts of computational resources, making them challenging to run on low-power devices such as the Raspberry Pi. One approach to addressing this challenge is to use pruning techniques to reduce the size of the deep learning models. Pruning involves removing unimportant weights and connections from the model, resulting in a smaller and more efficient model. Pruning can be done during training or after the model has been trained. Another approach is to optimize the deep learning models specifically for the Raspberry Pi architecture. This can include optimizing the model's architecture and parameters to take advantage of the Raspberry Pi's hardware capabilities, such as its CPU and GPU. Additionally, the model can be optimized for energy efficiency by minimizing the amount of computation required. Pruning and optimizing deep learning models for the Raspberry Pi can help overcome the computational and energy constraints of low-power devices, making it possible to run deep learning models on a wider range of devices. In the following sections, we will explore these approaches in more detail and discuss their effectiveness for optimizing deep learning models for the Raspberry Pi.

## I. INTRODUCTION

Running deep learning models [1] on low-power devices has become increasingly popular in recent years. However, low-power devices such as the Raspberry Pi face several challenges and limitations in terms of computational resources and energy efficiency. These limitations make it difficult to deploy deep learning models on low-power devices without significant optimization. One approach to optimizing deep learning models for low-power devices is pruning, which involves removing unimportant weights from the neural network. Pruning can significantly reduce the computational resources required to run a deep learning model, while maintaining or even improving its accuracy. Another approach is optimizing the neural network architecture, which involves changing the number and arrangement of the network's layers to reduce computational requirements while maintaining accuracy. In this paper, we focus on pruning and optimizing deep learning models for the Raspberry Pi, a popular low-power device. We investigate the impact of pruning on the computational performance and energy efficiency of deep learning models, and compare the performance of different pruning techniques. Additionally, we explore the benefits of optimizing the neural network architecture for the Raspberry Pi, and propose an optimized architecture for the device. Our experimental results demonstrate the effectiveness of these techniques in enabling the deployment of deep learning models on low-power devices.

The aim of the research was to study the effect of pruning and compression optimization techniques on deep image classification models when running real time inferences on a Raspberry Pi.

In the following sections, we will discuss the existing techniques for optimizing deep learning models on low-power devices, including pruning. Then, we will give the description of the experimental setup, including the Raspberry Pi hardware and software configuration. The next, we will provide the details on the dataset used and the deep learning model architecture and explanation of the pruning process and how it was applied to the deep learning model. Followed by, discussing the results and analysis. Finally, we will explain the interpretation of the results in light of the literature review, the implications of the findings for the practical deployment of deep learning models on low-power devices, summary of the key contributions and findings of the paper and the implications of the study for the field of deep learning on low-power devices.

## II. LITERATURE REVIEW

The following survey includes different architectures and their performances. The Raspberry Pi was used for animal detection in the wild [2] with the help of Raspberry Pi Model 3B+. They used a custom CNN which was trained on ImageNet [3] and VIPeR [4] datasets and retrained on snow leopard dataset to identify snow leopards in the wild. They were able to predict snow leopards with an accuracy of 0.97 for pre-downloaded images and 0.74 for live images. The inference time indicated in the research was 29.2 seconds for 347 images which is 84.15ms per image. The results were promising relative to the hardware cost and inference speed since it has the capacity to capture and identify live video. Benchmarking several edge devices which are used for inferencing of deep neural models are discussed in the DeepEdgeBench [5] . The DeepEdgeBench has benchmarked several leading devices (SoCs) including Nvidia's Jetson Nano, Asus Tinker Edge R, Google Coral Dev, Raspberry Pi and Arduino Nano microcontroller. The tests were performed using MobileNetV2, MobileNetV2 Lite, MobileNetV2 quantized and MobileNetV1 quantized models on 5000 images. The inference speed for an image and the Top-1 accuracy for the test images are tabulated as in Table 1.

Table 1: Results of DeepEdgeBench

| Model | MobileNetV2 | MobileNetV2 TF Lite | MobileNetV2 quantized | MobileNetV1 quantized |
|---|---|---|---|---|
| InferenceTime/image (ms) | 207.54 | 193.02 | 128 | 4.4 |
| Top 1 Accuracy | 0.733 | 0.733 | 0.726 | 0.364 |

A method to reduce latency and size is discussed in [6] where instead of a larger model a Neural Network Tree is used. Here smaller architectures are used to subclass inputs and classify them via multiple architectures until the input falls into a specific category. Latency, memory usage and power are reduced by this process and was tested on a Raspberry Pi 3 and Raspberry Pi Zero. The inference time comparison of different models using the Raspberry Pi 3 on the CIFAR-100 dataset are indicated on Table 2.

Table 2 Results obtained by [6]

| Model | VGG-Pruned | CondenseNet | MNN-Tree |
|---|---|---|---|
| Inference time per image (ms) | 934 | 7196 | 506 |

A simpler convolution neural network was used in tests done by [7] which used Raspberry Pi 3 to measure the inference time on 100 images belonging to cats and dogs. The model was a simple 5-layer CNN which was trained for 90 epochs. The inference time per image was recorded as 16.37 milli-second. A similar approach was taken by [8] which used 3 CNN models to classify the German Traffic Sign Recognition Benchmark dataset which comprise of 27 classes. Classification was done with an accuracy of 98.57% using the models and was tested on a Raspberry Pi 2B. The average time per image classification was 0.72 seconds. More promising results were found in [9] where a CNN model was used to classify images extracted from Vehicle Make and Model Recognition dataset which had 10 classes. Three models were trained using InceptionV3, VGG16 and MobileNet architectures. A Raspberry Pi 3B+ was used as the testing hardware with an Intel Neural Compute Stick which is hardware an accelerator for neural networks. A model optimizer was used to optimize the trained model to accelerate inference speed.

Table 3 Results obtained by [9]

| | InceptionV3 | VGG16 | MobileNet |
|---|---|---|---|
| Model load time (s) | 52 | 50 | 71 |

### III. METHODOLOGY

In this section, we will discuss the description of the experimental setup, including the Raspberry Pi hardware and software configuration, the details on the dataset used and the deep learning model architecture and the explanation of the pruning and optimization process and how it was applied to the deep learning model.

*A. Raspberry Pi*

There are several versions of Raspberry Pi released, the latest being the Raspberry Pi 4 Model B. Various peripheral devices can be connected to a Pi using the ports available. A key feature of the Raspberry Pi is that it contains general inputs and outputs (GPIO) which can be electrically connected to sensors or actuators and can be controlled according to a user program. The computer is powered by a 5V DC power supply and consumes less power than a modern computer.

The Raspberry Pi has the capability to run a variety of software including several operating systems. The widely used operating system is the official operating system of the Raspberry Pi Foundation which is the Raspberry Pi OS which is based on Debian Linux.

Software and APIs related to machine learning and deep learning such as Python, TensorFlow, OpenCV etc. can be installed on the Raspberry Pi and can be used as an edge device to run inferences. The model building and training is particularly difficult to be implemented on a Pi due to limited resources.

The Raspberry Pi should be configured to run deep learning models. Although the models are created and trained in a much more powerful system such as a PC, running the models on the Raspberry Pi requires pre-processing and loading the models which require native TensorFlow.

The details of the OS running on the Raspberry Pi 4 Model B used to run the models is given by.

- Model : Raspberry Pi 4 Model B Rev 1.5
- OS Name: Debian GNU/Linux, Version 11
- Version codename: bullseye
- Architecture: 64 bit
- CPU : Broadcom BCM2711, Quad core Cortex-A72 (ARM v8) 64-bit SoC @ 1.5GHz

The APIs and the modules installed will depend on the architecture and the OS of the Raspberry Pi. Before installing the APIs however, a virtual environment is created in the system so that that external package will not impact what we are going to test. The virtual environment will contain only the packages we install for our requirement and will prevent dependency issues from pre-installed packages.

*B. Datasets*

- MNIST and Svhn datasets: both datasets are divided to training and validation sets in the ratio of 70% to 30% respectively. Since the images are not complicated no augmentation is used. A custom CNN model was created using the inbuilt layers in TensorFlow.
- Horses or humans, cats_vs_dogs, and Flowers datasets: Unlike in the previous models transfer learning is used to train the model. This is done by using a pre-trained model which is VGG16, MobileNetV2 and ResNet 50 respectively.

*C. Optimization Techniques*

- Weight Pruning [10, 11]: We use the magnitude-based weight pruning technique to prune the weights with different sparsity levels in the range S = [0.25, 0.50, 0.60, 0.70, 0.80, 0.90, 0.95, 0.97, 0.99]. We then fine-tune the pruned model for a few epochs to recover the accuracy.
- Model Conversion: The conversion process is a set of tools to convert of the original models and the pruned model to TensorFlow Lite. TensorFlow Lite is a set of tools that allow machine learning models to run on mobile, embedded, and edge devices.

- Quantization[12]: We use post-training quantization technique to reduce the precision of the weights from float32 to int8. We then fine-tune the quantized model for a few epochs to recover the accuracy.
- TensorFlow Lite Delegates: Hardware acceleration of the TensorFlow models can be enables with the use of delegates which leverage on-device accelerators such as the GPU and processor. TensorFlow Lite utilizes CPU kernels that are optimized for ARM Neon instruction set, but the CPUs are multi-purpose and are not specifically optimized for Machine Learning models. Due to complications arising from different accelerator architectures, each one cannot execute every operation in a neural network. TensorFlow Lites' Delegate API bridges the TFLite runtime and lower APIs to solve this problem

### D. Benchmarking the Models

Benchmarking is the measure of relative performance of a computer program by running several tests on a specific system. Data or images for the model to infer is stored in folder alongside each of the models. For inferencing testing, a sample of 100 images for each dataset is used so a fair metric can be calculated. For each model the images are loaded one by one from the disk instead of batches and the time is calculated from the loading of the image to the final prediction of the class. The time take per each inference is stored and is used to find the average inference time. However, the inference time for the first image of the dataset is measured separately since during this run the loading of packages such as TensorFlow takes more time than the others and the rest of the inferences are much quicker. Hence, in equation the inferencing starts from the 2$^{nd}$ image rather than the first.

$$t_{infer\_1} = t_{end\_infer\_1} - t_{start\_infer\_1} \quad (1)$$
$$t_{infer\_i} = t_{end\_infer\_i} - t_{start\_infer\_i} \quad \text{for i = 2 to N} \quad (2)$$

Where,
$t_{end\_infer\_i}$ is the end time of the $i^{th}$ inference
$t_{start\_infer\_i}$ is the start time of the $i^{th}$ inference
$t_{infer\_i}$ is the time taken for the $i^{th}$ inference
N is the total number of test images

The average inference time $\bar{t}_{infer}$ can be calculated as:

$$\bar{t}_{infer} = \frac{t_{infer}}{N-1} \quad (3)$$

The standard deviation, $std\left(t_{infer}\right)$ can calculated as

$$std\left(t_{infer}\right) = \frac{\sum_{i=2}^{N}(t_{infer\_i} - \bar{t}_{infer})^2}{N-1} \quad (4)$$

The standard error, $ste\left(t_{infer}\right)$ is calculated as

$$ste\left(t_{infer}\right) = \frac{std\left(t_{infer}\right)}{\sqrt{N-1}} \quad (5)$$

The standard deviation gives an understanding of the distribution of the inference time. We can understand how far away the inferences are from the mean inference time. A low value indicates that the inference time of the images are very close to each other.

On the other hand, the standard error is the approximate standard deviation of a sample of statistical sample population. When the population is large the standard error tends to be smaller because standard error is inversely proportional to the population size.

The datasets for each model were kept in separate folders and the data belonging to different classes were divided into sub folders. The script was run using the Python interpreter in the Raspberry Pi.

### IV. RESULTS AND ANALYSIS

In this section we present the following:

- Presentation of the results obtained from the experiments
- Comparison of the performance of the pruned and unpruned models in terms of accuracy, speed, and memory usage
- Discussion of the trade-offs between accuracy and efficiency achieved through pruning

### A. Pruning the weights

Diagram 1 shows the comparison of the pruned model with the original model in terms of the number of parameters and accuracy/loss with different level of sparsity S on different datasets.

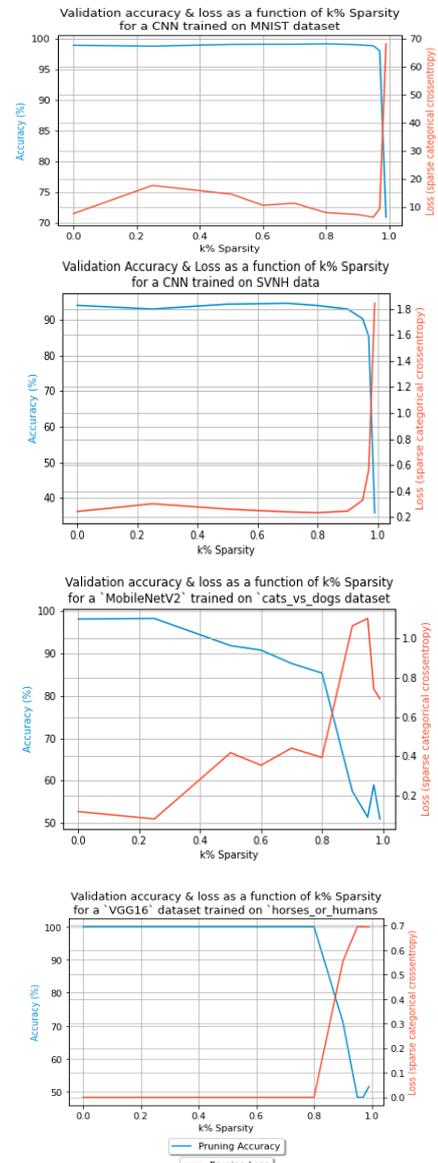

Diagram 1: Variation of validation accuracy & validation loss against different sparsity values of the model

## B. Pruning the Models for Best Accuracy

Once the models were pruned as in the previous section to observe the variation of accuracy and loss with different sparsity levels, we must select a sparsity level for each model to optimize further. One key observation is that the five models performed uniquely to pruning and the accuracy variation is unique to the model or even the dataset which was not tested. The five model were pruned again to a certain sparsity without affecting the accuracy significantly as in Table 5.

Table 5 Selected sparsity levels for further optimization of the models

| Model | Sparsity (k) | Epoch | Validation accuracy | |
|---|---|---|---|---|
| | | | before pruning(%) | after pruning(%) |
| MNIST | 0.7 | 5 | 98.97 | 100 |
| SVHN | 0.8 | 10 | 94.32 | 96.87 |
| Horses/humans | 0.8 | 20 | 99.67 | 100 |
| Cats vs dogs | 0.5 | 20 | 97.91 | 93.75 |
| Flowers | 0.7 | 20 | 92.55 | 90.62 |

When pruning and retraining the same compiling and fitting hyper-parameters were used changing only the number of epochs the model was trained for. The model accuracy of the MNIST, 'svhn' and 'horses or humans' datasets have increased after pruning. We can conclude that most of the parameters in the model has no impact on the model accuracy and therefore can be removed. However, the models for 'cats vs dogs' and 'flowers' data got a reduced accuracy than the original model after pruning. Next the models should be prepared to be used in the Raspberry Pi by converting them to a suitable format which will be discussed in the later.

## C. Benchmark Results of the Keras Models[1] and TensorFlow Lite

The Keras models were used to benchmark the performance of the Raspberry Pi and the results are shown in Table 6 and lite model in Table7.

Table 6 Inference results for test images using the Keras models

| Model / Time(ms) | MNIST | svhn | horses or humans | cats vs dogs | TF flowers |
|---|---|---|---|---|---|
| $t_{infer\_1}$ | 539.58 | 527.06 | 1861.52 | 2520.03 | 3886.11 |
| $\bar{t}_{infer}$ | 114.26 | 120.55 | 1386.21 | 241.58 | 1079.45 |
| $std\ (t_{infer})$ | 6.55 | 6.90 | 36.71 | 8.57 | 26.19 |
| $ste\ (t_{infer})$ | 0.66 | 0.70 | 2.61 | 0.87 | 2.65 |

Table 7 Inference results of the pruned / unpruned and (or) quantized TensorFlow Lite models

| Model / Time(ms) | MNIST | svhn | horses or humans | cats vs dogs | TF flowers |
|---|---|---|---|---|---|
| **Pruned TensorFlow Lite models** | | | | | |
| $t_{infer\_1}$ | 21.32 | 201.86 | 3648.91 | 123.36 | 1699.87 |
| $\bar{t}_{infer}$ | 3.73 | 9.11 | 3426.74 | 85.15 | 1580.09 |
| $std\ (t_{infer})$ | 0.26 | 0.05 | 4 | 1.91 | 3.24 |
| $ste\ (t_{infer})$ | 0.03 | 0.05 | 0.28 | 0.19 | 0.33 |
| **Unpruned TensorFlow Lite models** | | | | | |
| $t_{infer\_1}$ | 38.62 | 124.79 | 3559.50 | 38.30 | 1090.45 |
| $\bar{t}_{infer}$ | 3.74 | 84.57 | 3475.11 | 9.54 | 1009.91 |
| $std\ (t_{infer})$ | 0.24 | 2.07 | 8.86 | 0.75 | 8.51 |
| $ste\ (t_{infer})$ | 0.02 | 0.21 | 0.63 | 0.08 | 0.82 |
| **Pruned & quantized Lite models** | | | | | |
| $t_{infer\_1}$ | 28.90 | 155.24 | 1976.78 | 36.08 | 2006.02 |
| $\bar{t}_{infer}$ | 3.06 | 103.59 | 1765.95 | 7.21 | 1580.21 |
| $std\ (t_{infer})$ | 0.21 | 1.98 | 1.65 | 0.21 | 42.71 |
| $ste\ (t_{infer})$ | 0.02 | 0.20 | 0.12 | 0.02 | 4.31 |
| **Unpruned & quantized Lite models** | | | | | |
| $t_{infer\_1}$ | 29.87 | 156.41 | 1951.04 | 32.78 | 1832.24 |
| $\bar{t}_{infer}$ | 3.09 | 103.49 | 1768.14 | 7.21 | 1595.60 |
| $std\ (t_{infer})$ | 0.29 | 1.97 | 1.59 | 0.34 | 61.37 |
| $ste\ (t_{infer})$ | 0.03 | 0.20 | 0.11 | 0.03 | 6.20 |

We notice that the variation of the initial inference time for the MNIST and svhn models where the initial inference time is much lower for TF lite models as well as the quantized models. The reduction is nearly 10 times than that of the Keras model in both cases.

On the other hands, the initial inference time of the TF Lite and quantized 'cats vs dogs' models are reduced nearly by a factor of 25 times than that of the Keras model. For the TF flowers model the initial inference time is halved when the TF Lite and quantized models are used. The horses vs humans shows a different characteristic than the other models where the TF Lite initial image inference time is greater than the Keras model, however the quantized models have nearly equal latency to the Keras model.

When considering the average inference time of the models, the time taken per inference is less than the initial inference. The MNIST and svhn models show similar characteristics for the average inference time where the quantized models take up less time per image inference. The MNIST TF Lite model inference time has reduced by a factor of 30 times than that of the Keras model and for the svhn models the reduction is nearly 10 times. The TF flowers quantized models shows an increase in the inference time when compared to the Keras model. The cats vs dogs model has a reduced average inference time over the Keras model which has more than halved. The mean inference time of horses or humans model increases for the TF Lite models but decrease when the model is quantized.

## D. Accuracy variation of the models

Now we will look at the validation accuracy of the models after pruning and converting which is shown in Table 8.

---

[1] We use the term Keras models to distinguish to the lite models (TensorFlow Lite) that discuss earlier

Table 8 Model accuracy Keras, TF Lite & quantized pruned and unpruned models

| Model Accuracy (%) | MNIST | svhn | horses or humans | cats vs dogs | TF flowers |
|---|---|---|---|---|---|
| Keras model | 98.97 | 97.91 | 99.67 | 94.32 | 92.55 |
| Unpruned TFLite | 100.00 | 87.50 | 100.00 | 96.87 | 87.5 |
| Pruned TFLite | 100.00 | 87.50 | 100.00 | 96.87 | 87.5 |
| Unpruned quantized | 100.00 | 93.75 | 96.87 | 18.75 | 87.5 |
| Pruned quantized | 100.00 | 93.75 | 96.87 | 18.75 | 87.5 |

The MNIST model accuracy has increased when the models are converted to the TF Lite format and quantized. The 'cats vs dogs' model accuracy drops when converted to TF Lite format but increases when quantized, however the accuracy is slightly less than the Keras model. The 'horses or humans' model accuracy increase when converted to TF Lite model but reduces when quantized. For the 'svhn' model the accuracy increases when the model is converted to the TF Lite model but reduces drastically when the model is quantized. The TF flowers model accuracy remains the same when it is quantized or converted to TF Lite format and is slightly less than the Keras model.

*E. Performance using TensorFlow Lite Delegates:*

Arm NN [2] has Tflite Delegate library which is accelerating certain TensorFlow Lite operators on Arm hardware. The operations that are compatible with Arm NN delegate are executed and the ones which are not are off loaded to the default or TensorFlow Lite operations and executed on the CPU.

The Arm NN library should be installed separately on the Raspberry Pi to be activated so that it loads when inferences are made. The inferences on the test images done with the Arm NN delegate is shown in Table 9. Using a delegate has no effect on the MNIST model since the model is smaller and already makes fast inferences. There is a slight drop in initial image latency for the svhn model when the Arm NN delegate is in use. There is a considerable drop-in inference time for the 'cats vs dogs' model when the delegate is used but significant reduction in time is seen for the 'horses or humans' and 'TF flowers' models where it has nearly halved relative to the Keras model.

Similar effect can be seen on the average inference time when the delegate is used. A significant drop in the mean inference time can be seen for the horses or humans and TF flowers model when the delegate is used. When the pruned and unpruned models are converted to TF Lite format the inference time increased for the two models but due to the Arm NN delegate the inference time reduces and performs better than the Keras model.

Table 9 Inference results for test images using the unpruned and unpruned TF Lite models with Arm NN delegate

| Unpruned TF Lite models | | | | | |
|---|---|---|---|---|---|
| Model / Time (ms) | MNIST | svhn | horses or humans | cats vs dogs | flowers |
| $t_{infer\_1}$ | 31.99 | 95.40 | 596.41 | 39.06 | 561.15 |
| $\bar{t}_{infer}$ | 3.75 | 54.23 | 458.53 | 6.45 | 487.95 |
| $std\,(t_{infer})$ | 0.56 | 1.90 | 4.24 | 0.12 | 3.46 |
| $ste\,(t_{infer})$ | 0.06 | 0.19 | 0.30 | 0.01 | 0.35 |
| **Pruned TF Lite** | | | | | |
| $t_{infer\_1}$ | 31.05 | 92.60 | 594.59 | 38.62 | 554.80 |
| $\bar{t}_{infer}$ | 3.62 | 54.80 | 456.16 | 6.47 | 485.54 |
| $std\,(t_{infer})$ | 0.83 | 2.45 | 4.30 | 0.16 | 4.33 |
| $ste\,(t_{infer})$ | 0.08 | 0.25 | 0.31 | 0.02 | 0.44 |

V. CONCLUSION

From the results obtained, we can come to the following conclusions:
- Model accuracy does not change significantly when deep image classification models are pruned. In some cases, the model accuracy is increased due to pruning while in some cases there is a slight reduction. Custom CNN validation accuracy increases in both the cases but when pretrained layers are used the accuracy is reduced.
- When the models are converted to TF Lite format some models perform better than the Keras models. Custom CNN models have a lower inference time than their Keras models. However, models with the ResNet50 and VGG16 pretrained layers perform even worse than their Keras models. The inference time for the models which used these pretrained layers was increased. Since MobileNet was designed for edge devices the model which used this pretrained layer performed better than its Keras model.
- When the models are quantized (pruned and unpruned) the inference time is reduced for Custom CNNs which is even lower than the TF Lite versions. However, like the TF Lite versions the models based on ResNet50 and VGG16 shows an increase in the inference time. Model based on MobileNet performs even better than its Keras model but worse than the TF Lite version.
- The best inference results were obtained when the Arm NN delegate was used on the Raspberry Pi. All the models outperform earlier tests with regards to inference time and are reduced significantly. The Arm NN delegate can only be used for TF lite versions of the models and does not work on the quantized models.
- Model size does not change significantly when the models are converted to the TF Lite format when compared to the Keras format.
- Model size is not affected by pruning. The pruned and unpruned models have the same model size when they converted to the TF Lite version or quantized version.

---

[2] GitHub - ARM-software/armnn: Arm NN ML Software. The code here is a read-only mirror of https://review.mlplatform.org/admin/repos/ml/armnn

- Model accuracy is not affected by pruning but depends on the format. The pruned and unpruned TF Lite versions have the same accuracy. The quantized models have the same accuracy and does not depend on pruning.
- The best results in terms of accuracy and speed were obtained for TF Lite models with the Arm NN delegate on a Raspberry Pi.

REFERENCES

[1] S. Ameen and S. Vadera, "A convolutional neural network to classify American Sign Language fingerspelling from depth and colour images," *Expert Systems,* vol. 34, no. 3, p. e12197, 2017.

[2] B. H. Curtin and S. J. Matthews, "Deep learning for inexpensive image classification of wildlife on the Raspberry Pi," in *2019 IEEE 10th Annual Ubiquitous Computing, Electronics & Mobile Communication Conference (UEMCON)*, 2019: IEEE, pp. 0082-0087.

[3] J. Deng, W. Dong, R. Socher, L.-J. Li, K. Li, and L. Fei-Fei, "Imagenet: A large-scale hierarchical image database," in *2009 IEEE conference on computer vision and pattern recognition*, 2009: Ieee, pp. 248-255.

[4] D. Gray and H. Tao, "Viewpoint Invariant Pedestrian Recognition with an Ensemble of Localized Features," *ECCV (1),* vol. 2008, pp. 262-275, 2008.

[5] S. P. Baller, A. Jindal, M. Chadha, and M. Gerndt, "DeepEdgeBench: Benchmarking deep neural networks on edge devices," in *2021 IEEE International Conference on Cloud Engineering (IC2E)*, 2021: IEEE, pp. 20-30.

[6] A. Goel, S. Aghajanzadeh, C. Tung, S.-H. Chen, G. K. Thiruvathukal, and Y.-H. Lu, "Modular neural networks for low-power image classification on embedded devices," *ACM Transactions on Design Automation of Electronic Systems (TODAES),* vol. 26, no. 1, pp. 1-35, 2020.

[7] H. A. Shiddieqy, F. I. Hariadi, and T. Adiono, "Implementation of deep-learning based image classification on single board computer," in *2017 International Symposium on Electronics and Smart Devices (ISESD)*, 2017: IEEE, pp. 133-137.

[8] C. F. Silva and C. A. Siebra, "An investigation on the use of convolutional neural network for image classification in embedded systems," in *2017 IEEE Latin American Conference on Computational Intelligence (LA-CCI)*, 2017: IEEE, pp. 1-6.

[9] E. Kristiani, C.-T. Yang, and C.-Y. Huang, "iSEC: an optimized deep learning model for image classification on edge computing," *IEEE Access,* vol. 8, pp. 27267-27276, 2020.

[10] S. Han, J. Pool, J. Tran, and W. Dally, "Learning both weights and connections for efficient neural network," *Advances in neural information processing systems,* vol. 28, 2015.

[11] S. Vadera and S. Ameen, "Methods for pruning deep neural networks," *IEEE Access,* vol. 10, pp. 63280-63300, 2022.

[12] S. Han, H. Mao, and W. J. Dally, "Deep compression: Compressing deep neural networks with pruning, trained quantization and huffman coding," *arXiv preprint arXiv:1510.00149,* 2015.